\documentclass[prx,reprint,superscriptaddress,twocolumn,showkeys]{revtex4-1}
\usepackage[utf8]{inputenc} 
\usepackage{amsmath}
\usepackage{braket}
\usepackage{graphicx}
\usepackage{subfigure}
\usepackage{pgfplots}
\usepackage{csquotes}
\usepackage{hhline}
\usepackage{amssymb}

\usepackage{dcolumn}
\usepackage{tabularx}
\usepackage[colorlinks=true,linkcolor=blue,citecolor=blue,urlcolor=blue]{hyperref}
\usepackage{longtable}
\usepackage{braket}
\usepackage{float}
\usepackage{listings}
\usepackage{color}
\definecolor{mygreen}{rgb}{0,0.6,0}
\definecolor{mygray}{rgb}{0.5,0.5,0.5}
\definecolor{mymauve}{rgb}{0.58,0,0.82}

\newcolumntype{C}{>{\centering\arraybackslash}X}
\begin{document}

\title{Experimental demonstration of the violations of Mermin's and Svetlichny's inequalities for W- and GHZ-class of states}

\author{Manoranjan Swain}
\email{swainmanoranjan333@gmail.com}
\author{Amit Rai}
\email{amitrai007@gmail.com}
\affiliation{National Institute of Technology, Rourkela, 769008, Odisha, India}

\author{Bikash K. Behera}
\email{bkb13ms061@iiserkol.ac.in}
\author{Prasanta K. Panigrahi}
\email{pprasanta@iiserkol.ac.in}
\affiliation{Department of Physical Sciences,\\ Indian Institute of Science Education and Research Kolkata, Mohanpur 741246, West Bengal, India}

\begin{abstract}
Violation of Mermin's and Svetlichny's inequalities can rule out the predictions of local hidden variable theory and can confirm the existence of true nonlocal correlation for n-particle pure quantum systems (n$\geq$3). Here we demonstrate the experimental violation of the above inequalities for W- and GHZ-class of states. We use IBM's five-qubit quantum computer for experimental implementation of these states and illustration of inequalities' violations. Our results clearly show the violations of both Mermin's and Svetlichny's inequalities for W and GHZ states respectively. Being a superconducting qubit-based quantum computer, the platform used here opens up the opportunity to explore multipartite inequalities which is beyond the reach of other existing technologies.
\end{abstract}

\maketitle

\section{Introduction}

Both Bell's inequality~\cite{qv_f1} and the CHSH inequality~\cite{qv_f18} were formulated for two particles, to distinguish between non local hidden variable theory and quantum mechanics. Clauser and Stuart~\cite{qv_f19} and Aspect $\emph{et al.}$~\cite{qv_f20} through experimental Bell tests proved the predictions of quantum mechanics to be true using entangled photons. With the advancement of optical systems, Hesen $\emph{et al.}$ reported loopholes free experimental test of Bell inequalities~\cite{qv_f6}. Recent development in the theoretical and experimental study of Bell's theorem can be found from the review article of Brunner \emph{et al.}~\cite{qv_f7}.

Mermin inequalities~\cite{qv_f2} are the extended Bell type inequalities derived for n particles (n $\geq3$), to test nonlocal quantum correlations between the entangled particles. Violation of these Bell type inequalities can characterize maximally entangled states~\cite{qv_f5}. However, the fact that violation of these can confirm true nonseparability of the particles is limited to only two particle case. Hence for a multi-particle system (n $\geq3$) maximal violation of Mermin's inequality alone cannot predict existence of true nonlocal correlation between the particles. Instead one can think of a model where violation of these inequalities can successfully explain m particle nonlocal correlation (m $\textless$ n) and local correlation with the rest for a n-particle state. This was first suggested by Svetlichny~\cite{qv_f4}. Later Svetlichny extended his idea in the form of an inequality for three particle case, violation of which not only confirms the existence of three particle entanglement but also true for three particle nonlocality, which has been generalized to n particles~\cite{qv_f9,qv_f10}.

For a particular case of three particles, there are two important class states, GHZ and W-state. These two states are non convertible to each other under local operations and classical communications~\cite{qv_f16}. The maximally entangled GHZ state~\cite{qv_f21,qv_f22} has been used in quantum key distribution~\cite{qv_f24}, quantum cheque demonstration~\cite{qv_f12}, superdense coding~\cite{qv_f25} etc. On the other hand the W-state~\cite{qv_f16} is found to be robust in maintaining entanglement against particle loss and its applications are seen in quantum teleportation~\cite{qv_f26}, quantum secure communication~\cite{qv_f27}, and quantum cloning machines~\cite{qv_f28} etc. Hence study of quantum correlation for these states plays a key role in quantum information theory. Alsina and Latorre have reported the violation of Mermin's inequality for different classes of GHZ states~\cite{qv_f3} using IBM's cloud computing platform. Experimental violation of Svetlichny's inequality has been explicated for three-photon GHZ states using photonic system \cite{qv_LavoieNJP2009}. Quantum correlations such as quantum contextuality, Leggett–Garg temporal correlations, and quantum discord have been studied in NMR Systems \cite{qv_MaheshLGQCA2017}. However, investigation of these inequalities for GHZ- and W-class of states in the current technology of superconducting quantum chip explores a new opportunity to experimentally realize. Here we show the violation of Svetlichny's inequality (SI) for a three- qubit GHZ state along with the violation of Mermin's inequality (MI) for three-qubit W-class state using IBM's quantum computer. The advantage of using IBM's superconducting qubit-based quantum computer is that other multipartite inequalities can be easily tested and verified which is hard and challenging to achieve using other existing technologies.

IBM Q is a superconducting qubit based operating system which offers open global access to a wide class of researchers and has found significant applications in an user friendly interface~\cite{qv_f8}. Violation of CHSH inequality~\cite{qv_f8} and Mermin's inequality for GHZ state~\cite{qv_f3}, experimental realization of quantum cheque~\cite{qv_f12}, demonstrations of non-Abelian braiding of Majorana modes~\cite{qv_f13}, verification of quantum algorithms \cite{qv_GangopadhyayQIP2018,qv_ZhukovQIP2018}, testing error correction codes \cite{qv_GhoshQIP2018,qv_SatyajitQIP2018}, building of a quantum repeater~\cite{qv_f11}, quantum simulation \cite{qv_LiertaarXiv2018,qv_HegadearXiv2018}, fidelity improvement \cite{qv_PokharelarXiv2018} are the examples of a few tasks which have been implemented successfully.

The rest of the paper is organized as follows. Section \ref{qv_SecII} discusses violation of Mermin's inequality for W-state and illustrates its experimental realization. Section \ref{qv_SecIII} explicates the violation of Svetlichny's inequality for GHZ state with experimental demonstration. Results are then presented and discussed in Section \ref{qv_SecIV}. We finally conclude in Section \ref{qv_SecV} providing future directions of the present work.

\section{Violation of Mermin's inequality for W-state \label{qv_SecII}} 
\begin{figure}[]
  \subfigure[]{\includegraphics[width=1.05\linewidth]{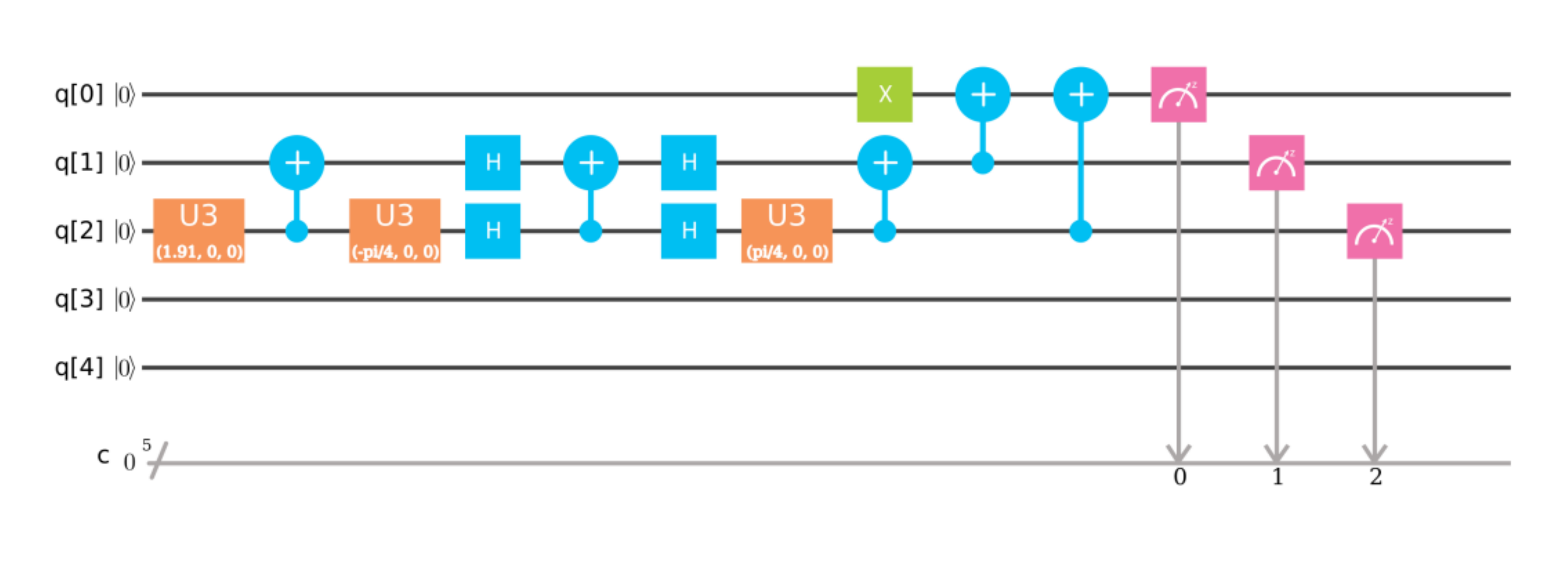}}
  \subfigure[]{\includegraphics[width=1.05\linewidth]{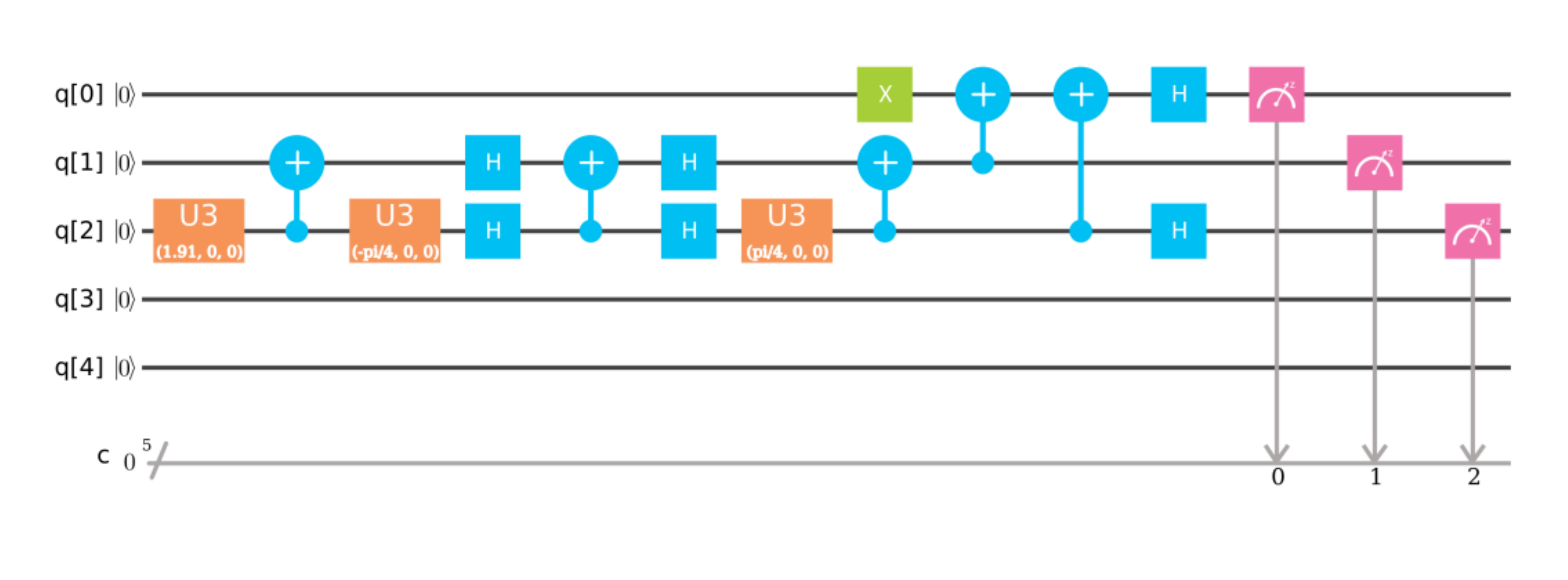}}
  \caption{(a) Circuit for implementation of maximally entangled W-state represented in Eq. \eqref{qv_Eq2}. (b) Execution of the measurement for the expectation value of $E(A'BC')$ on W-state.}
  \label{qv_Fig1}
\end{figure}

One of the Mermin's inequality for three particles is of the following form~\cite{qv_f14},

\begin{equation}\label{qv_Eq1}
\begin{aligned}
|M|=|E(ABC)-E(A B'C')-E(A'B'C)\\
-E(A'BC')|\leq 2
\end{aligned}
\end{equation}

where $A(A')$, $B(B')$ and $C(C')$ are arbitrary possible choice of measurements on particles 1, 2 and 3 respectively. The outcomes of all these measurements can either be +1 or -1. Our verification process is based on the calculation of correlation functions, e.g. $E(ABC)$. The observable $E(ABC)$ represents the expectation value of joint probability of outcomes of the measurements $A$, $B$, and $C$ on particles 1, 2 and 3 respectively. Calculation of all such terms appearing in Eq. \eqref{qv_Eq1} leads to the calculation of Mermin's polynomial. The value of which is then compared to the value predicted with local realism. Here, we show the violation of the above mentioned Mermin's inequality using a maximally entangled W-state. The W-state we have considered, is represented in Eq. \eqref{qv_Eq2} and generation of the same is shown in Fig. \ref{qv_Fig1} (a).

\begin{equation} \label{qv_Eq2}
\arrowvert W\rangle= \frac{1}{\sqrt{3}}(\arrowvert100\rangle+ \arrowvert010\rangle+ \arrowvert001\rangle)
\end{equation}

In the domain of quantum mechanics, these measurements can be regarded as spin measurements and are specified by linear combination of Pauli spin operators. By restricting the choice of spin measurements to XZ plane only, the quantum mechanical expectation value of the first term of the R.H.S of Eq. \eqref{qv_Eq1} can be calculated for the W state as,

\begin{equation}\label{qv_Eq3}
\begin{aligned}
E_{W}(ABC)=\langle W|\sigma (\hat{n_1})\otimes\sigma (\hat{n_2})\otimes\sigma (\hat{n_3})|W\rangle\\
 =-\frac{2}{3} cos(\theta_1+\theta_2+\theta_3)-\frac{1}{3}cos\theta_1 cos\theta_2 cos\theta_3,
\end{aligned}
\end{equation}

where $\theta_i(\theta'_i)$ are polar angles specifying the measurement direction $\hat{n}_i(\hat{n}'_i)$. Following the above expression, other terms of the inequalities can also be calculated. For the choice of measurements $\theta_1=\theta_2=\theta_3=\frac{1}{3}n\pi$ (n=0, $\pm$1, $\pm$2,...) and $\theta'_i$=$\theta_i$+$\pi$/2, the value of Mermin's polynomial ($|M|$) represented in Eq. \eqref{qv_Eq1} is calculated to be 3. This value is more than the classical bound of Mermin's polynomial predicted by local hidden variable (LHV) theory.\\

To execute this task in IBM QE, we set the polar angles $\theta_1=\theta_2=\theta_3$=0 and $\theta_1'=\theta_2'=\theta_3'=\pi/2$ for which the measurement direction for the unprimed and the primed ones become Z and X respectively. The measurements along Z and X are similar to previously known $\sigma_z$ and $\sigma_x$ measurements in IBM QE platform respectively. For convenience, the implementation of one of the measurements ($E(A'BC')$) is shown in Fig. \ref{qv_Fig1} (b).

Due to symmetry of the terms used in the inequality ($E(A B'C')$, $E(A'B'C)$ and $E(A'BC')$) under particle exchange, only one experiment on behalf of three has been executed as all of them would have obtained similar results. For calculation of  Mermin's polynomial experiments related to two different measurements ($ABC$ and $A'BC'$) are run, each for 1024 times. The output of the measurements are listed in Table \ref{qv_TabI}.

\begin{table}\label{qv_TabI}
 \centering
  \caption{Outcome of the results with their corresponding probabilities of $ABC$ and $A'BC'$ measurements. These probabilities are then converted to expectation values with which the Mermin's polynomial can be calculated as, $|M|=|E(ABC)-3 E(A'BC')|.$}
 \begin{tabular}{p{2.5cm} p{2.2cm} p{2.2cm}}
 \multicolumn{3}{c}{} \\
 \hline
 \hline
 Outcomes & Probabilities for $ABC$ &Probabilities for $A'BC'$\\
 \hline
 \textbf{000}   & 0.077    &0.363\\
 001   & 0.286  & 0.030\\
 010 &0.268 & 0.074\\
 \textbf{011}    &0.015 & 0.081\\
 100  &   0.286  & 0.065\\
 \textbf{101}  & 0.021  & 0.240   \\
 \textbf{110} & 0.008  & 0.096\\
 111 & 0.039  & 0.050\\
 \hline\hline
 \end{tabular}
\end{table}

Further proceeding to the calculation of Mermin's polynomial these probability outcomes need to be translated in terms of expectation values. This can be done in arranging the result in groups according to the parity of the number 1 (which has a value -1). The expectation values are then calculated by adding all probabilities of even parities (i.e. 000, 011, 101, 110) and subtracting the results of odd parities from them. With the above shown result the values of $E(ABC)$ and $E(A'BC')$ are calculated to be -0.758 and 0.561 respectively. Hence value of the polynomial represented in Eq. \eqref{qv_Eq1} is calculated to be 2.441. This shows the clear violation of the classical bound predicted by LHV theory for Mermin's inequality. 

\begin{table*}[t]\label{qv_TabII}
 \centering
  \caption{Outcome of the result and their corresponding probabilities of all sets of measurements on GHZ state.}
 \begin{tabular}{p{1.5cm} p{1.5cm}p{1.5cm}p{1.5cm}p{1.5cm}p{1.5cm}p{1.5cm}p{1.5cm}p{0.9cm}}
 \multicolumn{9}{c}{} \\
 \hline
 \hline
 Outcomes   &  $ABC$    &$ABC'$   &$AB'C$ &$A'BC$ &$A'B'C$ &$A'B'C'$ &$A'BC'$ &$AB'C'$
 \\ \hline
 \textbf{000}   & 0.079    & 0.082   &0.093  &0.085  &0.208   &0.188    &0.213   &0.248\\
         001    & 0.175    & 0.188   &0.214  &0.159  &0.068   &0.066    &0.052   &0.063\\
         010    &0.208     & 0.207   &0.174  &0.174  &0.062   &0.063    &0.046   &0.064\\
 \textbf{011}   &0.051     & 0.032   &0.036  &0.033  &0.153   &0.139    &0.151   &0.159\\
         100    &0.216     & 0.192   &0.193  &0.223  &0.080   &0.104    &0.089   &0.074\\
 \textbf{101}   &0.054     &0.056    &0.058  &0.062  &0.204   &0.190    &0.202   &0.168\\
 \textbf{110}   &0.040     &0.066    &0.053  &0.082  &0.182   &0.218    &0.212   &0.189\\
         111    &0.178     &0.176    &0.180  &0.183  &0.043   &0.031    &0.035   &0.033\\
 \hline    
        Total   &-0.553    &-0.527   &-0.521 &-0.477 &0.494   &0.471    &0.556   &0.530\\
 \hline\hline
 \end{tabular}
\end{table*}

\section{Violation of Svetlichny's inequality for GHZ state \label{qv_SecIII}}

\begin{figure}[]
\subfigure[]{\includegraphics[width=1.05\linewidth]{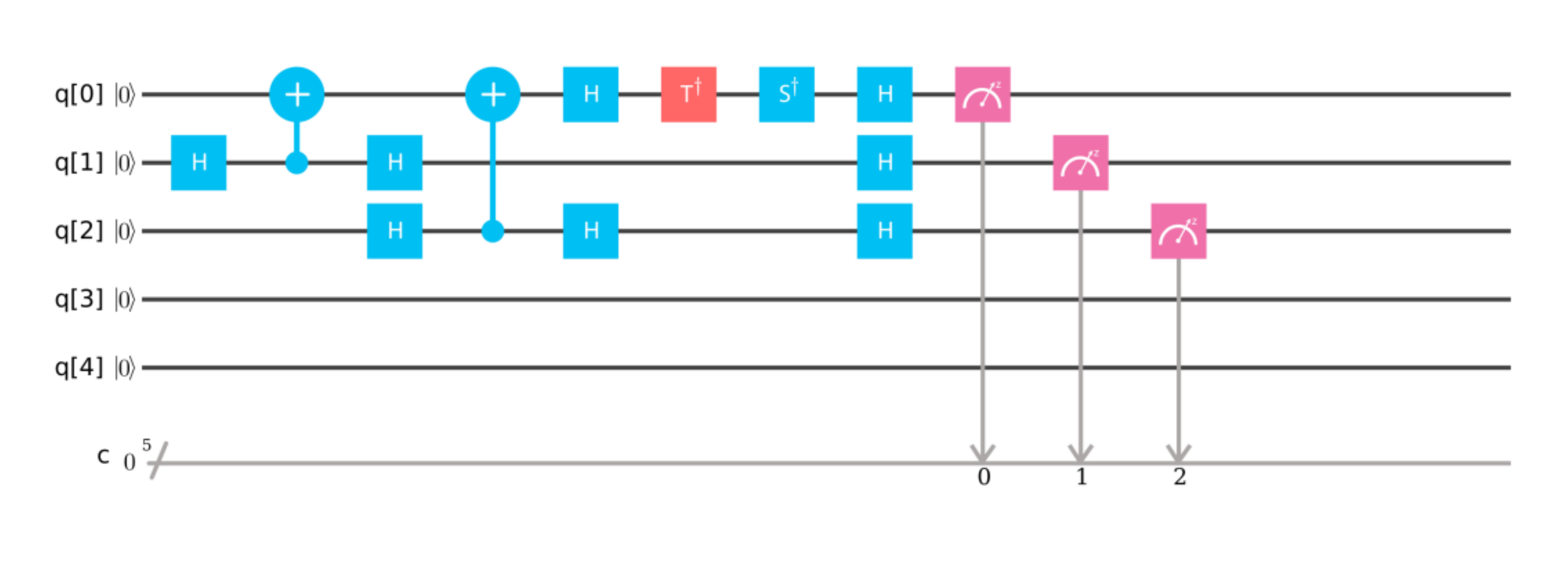}}
\subfigure[]{\includegraphics[width=1.05\linewidth]{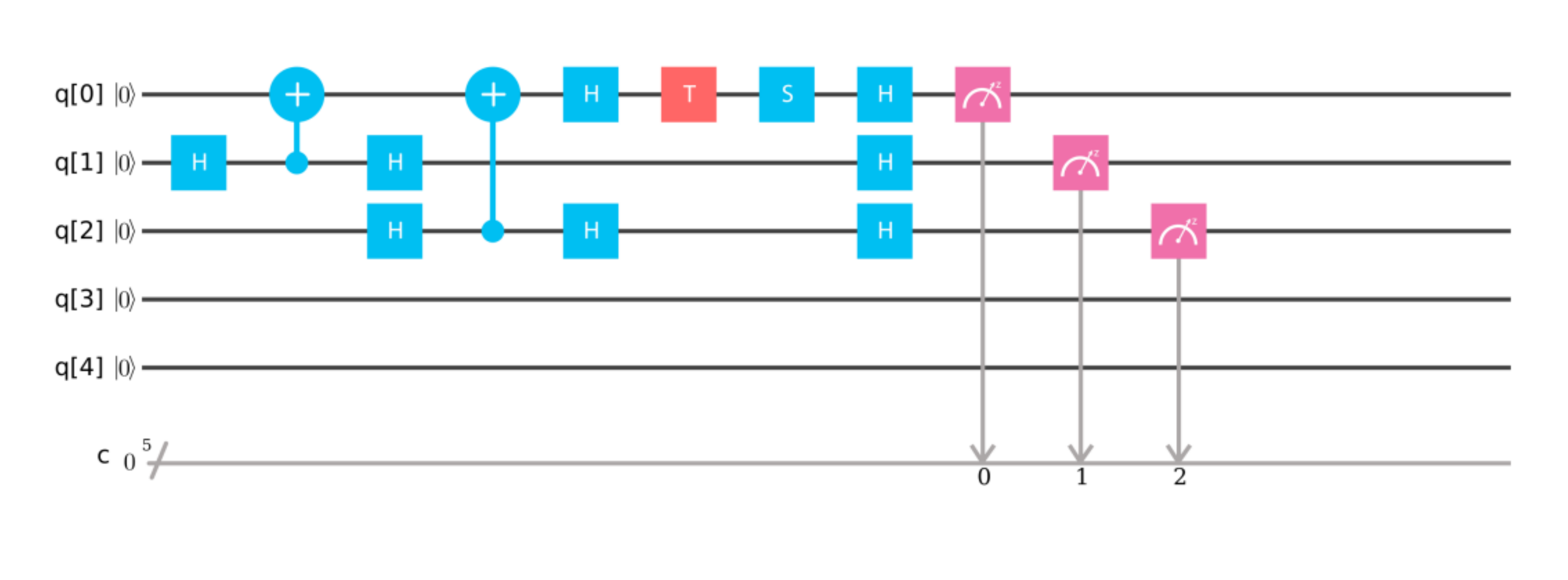}}
\caption{Circuit implementation for (a) $ABC$ and (b) $ABC'$ measurements on GHZ state respectively.}
\label{qv_Fig2}
\end{figure}

The Svetlichny's inequality for three particles~\cite{qv_f14} is represented below in Eq. \eqref{qv_Eq4}. This inequality can be treated similar to the extended Bell type inequalities which would contain terms of both $M$ and $M'$. $M'$ is another form of Mermin's inequality which can be obtained by taking primes of terms presented in Eq. \eqref{qv_Eq1}. For three particles there exist particular set of measurements for which MIs are maximally violated. However, these measurements may not violate SI i.e. such measurement settings may be used to check two particle nonlocality but three particle nonlocality tests may not be verified~\cite{qv_f14}. Our focus is on particular set of measurements using which SI can be violated. If violation of SI occurs, existence of both three particle entanglement and three particle nonlocality can be verified. We have considered here the case of three-qubit maximally entangled GHZ state (represented in Eq. \eqref{qv_Eq5}) to do experiment with, whose generation along with measurements is shown in Fig. \ref{qv_Fig2}.

\begin{equation}\label{qv_Eq4}
\begin{aligned} 
   |{S_v}|=|E(ABC)+E(ABC')+E(AB' C)+E(A' B C)\\
   -E(A' B' C')-E(A' B' C)-E(A' B C')-E(A B' C')|\leq 4
\end{aligned}
\end{equation}

\begin{equation}\label{qv_Eq5}
 |GHZ\rangle=\frac{1}{\sqrt{2}}(|000\rangle+|111\rangle)
\end{equation}

Mapping the measurements to linear combination of Pauli matrices and restricting the spin measurements in XY plane only, the calculated expectation value of E(ABC) is given by,

\begin{equation}\label{qv_Eq6}
\begin{aligned}
 E_{GHZ}(ABC)=\langle GHZ|\sigma (\hat{n_1})\otimes\sigma (\hat{n_2})\otimes\sigma (\hat{n_3})|GHZ\rangle\\
     =cos(\phi_1+\phi_2+\phi_3),
\end{aligned}
\end{equation}

where $\phi_i$ are the azimuthal angles specifying different direction of measurements in XY plane. For the choices $\phi_1+\phi_2+\phi_3$=(n+$\frac{3}{4}\pi$), where (n=0,$\pm$1,$\pm$2,...) and $\phi'_i$= $\phi_i$+$\pi$/2, the value of Svetlichny operator is algebraically calculated to be 4$\sqrt{2}$~\cite{qv_f14}. This is the maximum value of the operator predicted by quantum mechanics whereas according to local realism $|S_v|\leq$4.

To execute this task in IBM's five-qubit quantum chip `ibmqx4', we consider $\phi_1$=$\phi_2$=0, $\phi_3$= $\frac{3}{4}\pi$, $\phi_1'$=$\phi_2'=\pi/2$ and $\phi_3'=\frac{5}{4}\pi$. These values set the direction of measurements $ A, B, A'$ and $ B'$ as X, X, Y and Y direction respectively. Measurement in these directions are same as taking $\sigma_x$, $\sigma_x, \sigma_y$ and  $\sigma_y$ measurements in IBM quantum computer respectively. Similarly with the values of $\phi_3$ and $\phi_3'$, the measurement directions $C, C'$ becomes $(Y-X)/\sqrt{2}$ and -$(X+Y)/\sqrt{2}$ respectively. To measure spins in these directions we have to use additional single qubit gates T, T$^\dagger$, S and S$^\dagger$. For convenience, the measurement schemes for $ABC$ and $ABC'$ are shown in Fig. \ref{qv_Fig2}.

For eight different terms appearing in Eq. \eqref{qv_Eq4}, eight different measurements are run in ibmqx4. Each measurement was executed 1024 times. The outcome of the measurements are presented in Table \ref{qv_TabII}. Following the same procedure as described in Section \ref{qv_SecII} for Mermin's inequality, the probabilities are translated to expectation values. The Svetlichny's operator with the values obtained from experimental results, is calculated to be 4.129. As local realism sets a bound of 4 for this operator, our experimental result clearly shows the violation of Svetlichny's inequality.

\section{Results and discussion \label{qv_SecIV}}

The three particle states fall into two distinct categories, GHZ and W-state~\cite{qv_f16} from the perspective of teleportation. GHZ state, unitarily connected with the modified W-state shows stronger nonlocality than Bell states~\cite{qv_f17}. It therefore imperative to test the nature of quantum correlation of these two distinct cases. Violation of Mermin's inequality is regarded as a test of nonseparability of particles for a pure quantum state. Alsina and Latorre have shown violation of Mermin's inequality for three-, four- and five-qubit GHZ states~\cite{qv_f3}. Here we have shown violation of Mermin's inequality for a three-qubit W-state. For three particles, violation of only Mermin's inequality do not provide the true test for non-locality. However, violation of Svetlichny operator can strongly define the non-locality between the three entangled particles~\cite{qv_f4}. Our experimental result for three-qubit GHZ state obtains a value of 4.129 for Svetlichny's operator. This value is more than the value ($|S_v|\leq4$) predicted by local realism for this operator. Hence our experimental result is sufficient to show the violation of Svetlichny's inequality. A comparison between local realism and quantum mechanical predicted values of inequalities with experimentally obtained values for three particles is shown in Table \ref{qv_TabIII} . Also one important thing to consider, the platform we are using is not free from technical issues which limits us to obtain the ideal value of the experiment. Such issues include decoherence, gate errors and readout errors. For two particles, the experimental results of CHSH inequality \cite{qv_f8}, obtained an error about 0.03. However as in our case the number of particles as well as the number of measurements are more, the expected error is also more. These results can further be improved with minimization of decoherence effect, the gate and readout errors.\\

\begin{table}[h]
 \centering
 \caption{Comparison of local realism (LR) and quantum mechanics (QM) predicted values of Mermin inequality and Svetlichny inequality with experimentally obtained results for three particle states.}
 \begin{tabular}{p{2.1cm} p{2.1cm} p{2.1cm} p{2.1cm}}
 \multicolumn{3}{c}{} \\
 \hline\hline
Inequalities & LR  & QM & Experiment\\
 \hline
 MI   & 2    &4&2.441\\
 SI   & 4  & $4\sqrt{2}$ &4.129\\
 \hline\hline
 \end{tabular}
\label{qv_TabIII} 
\end{table}

\section{conclusion \label{qv_SecV}}

In conclusion we have shown the violations of Mermin's and Svetlichny's inequality for three-qubit W-state and GHZ states respectively using IBM's cloud computing platform. Our experimental results clearly show the violation to that of the values predicted by local realism for these inequalities. The SI test can further be performed for more number of qubits as a test of multi-particle nonlocal correlations.

\subsection*{Acknowledgments}
We acknowledge IBM team for providing free access to their cloud computing platform. The work presented here is only of the authors and do not include any technical contribution of IBM team.

\end{document}